# Characterization of Argon Plasma in a variable Multi-pole line Cusp Magnetic Field (VMMF) Configuration

A. D. Patel[*], M. Sharma, and, N. Ramasubramanian

*Institute for Plasma Research, HBNI, Bhat, Gandhinagar, Gujarat - 382428, India*

**\*E-mail:** *amitpatel@ipr.res.in*

## Abstract

This paper demonstrates a detailed characterization of argon plasma in a variable multi-pole line cusp magnetic field (VMMF). The VMMF has been produced by placing six electromagnets (with embedded profiled vacoflux-50 core) over a large cylindrical volume (1 m axial length and 40 cm diameter). The magnetic field have been measured by hall probe method and compared with simulated magnetic field by performing simulation using FEMM tools. Results from magnetic field simulation indicate that the rate of change of pole magnetic field (maximum magnetic field) with respect to magnet current for vacoflux-50 core is high (7.53 G/A) as compared to the simple air core electromagnet (2.15 G/A). The area of the nearly field free region (null region) in the chamber volume can be controlled without changing a number of pole magnets. From the experimental results, it has been observed that in this field configuration the confinement of the primary electrons increases and leak width of plasma decreases with increasing the magnetic field. Thus the mean density, particle confinement time and the stability of the plasma increase with increasing magnetic field. In addition to this, it has been found that the radial uniformity of the plasma density explicitly depends on the VMMF. It is also shown that the VMMF controls the scavenging of confined primary electrons and confinement of primary electron increased with magnetic field which helps to boost up the plasma density.

## I. Introduction

The utilization of multi-cusp magnetic field (MMF) has found a wide application in the development of ion sources due to its ability to confine a large volume of high-density uniform and quiescent plasma [1]. In fact the principle of plasma confinement by using MMF is mainly originated with fusion research [2-3], but it has been demonstrated that MMF based negative ion sources are capable for producing high current low emittance and stable ion beams which are essential for the new generation particle acceleration experiments. Actually to heat the core region of Tokomaks plasma up to the ignition temperatures, high power neutral beams having energy of the order of hundreds of keV (up to1MeV for $D^0$ or up to 870keV for $H^0$) are required [4]. For this application, the plasma source should be capable of producing quiescent, uniform and dense plasmas in order to produce a well-collimated high current density beam. As for example, the noise level and spatial density should be less than 10% over a several hundreds of square centimetres at a plasma density of a few times $10^{12} cm^{-3}$. It is found that MMF can confine large volume quiescent plasma at a density in the order of $10^{12} cm^{-3}$ [5].Thus MMF based plasma sources have achieved a great deal of attention to NBI (Neutral Beam Injection) systems. In addition to the above plasma (ion) sources confined by MMF are widely employed in a number of systems which include particle sources, etching [6], implantation [7], deposition [8-9] and fusion devices [4, 10].

MMF is usually created by placing rows of magnets along the length of cylindrical plasma chamber with alternating polarity. In 1973, Limpaecher *et. al*. first observed that MMF generated by permanent magnets can increase the density of a DC discharge plasma [1]. Later confinement of plasma in different cusp geometries has been investigated by Leung *et. al.* and it was observed that a full-line cusp geometry gives the highest density by confining primary electrons [11]. Actually there are three important parameters associated with MMF design for plasma sources; firstly, the area of field free region and secondly line cusp loss width of plasma and lastly control over the radial uniformity across the magnetic field. Here the field-free central region plays a crucial role in producing an well collimated ion source as it determines the volume available for the formation of uniform plasma as well as the working area to place the filaments. For proper emission of an electron from the filaments and hence to generate a uniformly dense plasma, the filament should be kept at the field free region. The area of the field-free region of MMF is usually adjusted by varying of a number of the pole magnets [10]. The second crucial confinement property involved with the cusp magnetic field is the effective plasma leak width i.e. the width of the profile of plasma escaping through the cusp [12-13]. The cusp loss width governs the efficiency of plasma source [12-13]. An exact scaling for the amount of plasma leaking is not known yet explicitly, though many scales ranging from ion gyro-radius to electron gyro-radius have been published in the literature [12-19]. The last crucial confinement property of MMF based plasma sources is radial uniformity of plasma density across the magnetic field. In earlier cusp magnetic field the radial uniformity changed with changing permanent magnets [20, 21].To boost up plasma density with increased radial uniformity of the plasma density across the magnetic field are the essential properties of all MMF based plasma sources.

Most of the earlier studies in the MMF configuration for plasma sources employed permanent magnet [1, 4-6, 9, 11, 20-21] and arrangement of magnets either in line cusp or in chequer board configuration. Also, the area of the field-free region can be changed only by varying the number of the poles (magnets). The size of the field free region is increased by increasing number of poles but the plasma loss area increases simultaneously with increasing the number of poles. The earlier cusp magnetic field devices have no control over the area of null region of cusp magnetic field and hence on the mean plasma density, radial uniformity of plasma density unless the number of poles are changed. The present device has the capability to vary the field free region without changing the number of poles by varying the magnetic field.

In this present article, we have demonstrated a detailed characterization of argon plasma in a variable multi-pole line cusp magnetic field (VMMF). The VMMF has been produced by placing six electromagnets (with embedded profiled vacoflux-50 core) over a large volume (1 m axial length and 40cm diameter). The edge of vacoflux-50 core material is profiled to avoid edge effect on magnetic field lines. Similar field lines have been observed by performing simulation using FEMM tools. The VMMF has control over the size of the null region without changing the number of the poles. Also, the rate of change of pole magnetic field (maximum magnetic field) with respect to magnet current for vacoflux-50 core electromagnet is too high compared to simple air core electromagnets. The argon plasma has been produced by a hot cathode filament based source located at the centre of one end of the chamber. The plasma is characterized for different values of a magnetic field of a VMMF and the results show that VMMF has control over mean plasma density, radial uniformity of plasma density and filtering of primary electron across the magnetic field.

Also, the proposed device is aimed to designed and fabricated for controlled study of plasma sources, Hall plasma thrusters, turbulence present at the cusp region and its effect on null region, as well as active experiment on quiescent plasma. Before embarking of proposed experiments, it is desired to characterize the plasma confined in a variable multi-pole line cusp magnetic field as well as magnetic field scaling with magnet current. The rest of the paper is arranged as follows: Sec. II describes the brief description of the experimental setup. Sec. III describes detail analysis on a variable cusp magnetic field simulation and comparison with the measured value. In addition, we have also described the profiling of magnetic field lines using profiled vacolfux-50 as a core material. We have also compare the simulation results for magnetic field values by changing core material i.e. air core and vacoflux-50 core. Sec. IV describes the experimental observation and results on confinement of primary electron and leak width of plasma and its effect on plasma parameter in a VMMF field will be discussed and conclude in the last section.

## II. Experimental Setup

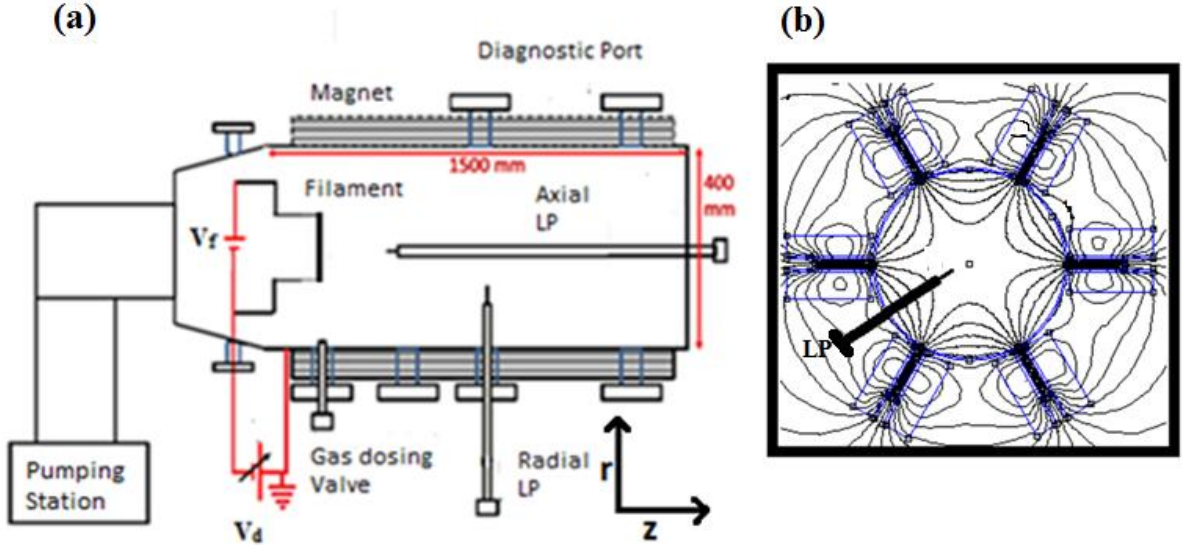

Figure 1. Schematic diagram of (a) experimental setup and (b) chamber cross sectional view of the multi-line cusp magnetic field plasma device, where LP-Langmuir Probe, $V_f$ - floating Power supply for filament heating, and $V_d$ - discharge power supply.

Figure 1, (a) shows a schematic diagram of experimental setup and (b) shows cross sectional view of experimental setup of the variable multi cusp magnetic field plasma device (MPD). The experimental setup consists of a cylindrical vacuum chamber (diameter = 40 cm and length = 1.5m). This chamber is pumped out by a combination of Rotary-TMP (430 l/s) pump capable of $10^{-6}$ mbar base pressure. The cusp magnetic field has been produced by six rectangular electromagnets and magnetic field is profiled using vacoflux-50 core material. The filamentary argon discharge plasma is produced using hot filament based cathode source. The plasma source (cathode) is two dimensional (8cm x 8cm) vertical arrays of five tungsten filaments; each filament has 0.5mm diameter and 8cm length. It is fitted from the conical reducer such that the filaments are inside the main chamber itself where the magnetic field is low. Also it has been taken care to push the source well inside the main chamber to avoid the edge effects of the magnets. These filaments are powered in parallel by a 500A, 15V floating power supply ($V_f$) while it is normally operated at around 16-19A per filament. The chamber was filled with rgon gas through a needle valve to a pressure $2 \times 10^{-4}$ mbar. The filament is biased with a voltage of -76 V with respect to the grounded chamber walls using discharge power supply ($V_d$). The primary electrons emitted from the filaments travel in the electrical field directions towards chamber wall anode, while they are confined by the cusp magnetic field lines. All experimental measurements are carried out at mid $(r, \theta)$ plane of the device which is z = 65 cm away from the filaments until and unless specified.

## III. A variable Multi-pole line cusp magnetic field (MMF)

Initially, A Finite Element Method Magnetics (FEMM) [22] simulation is performed for MMF assuming permanent magnets to finalize the size and axial length of the chamber to satisfy the basic experimental requirements. After a lot of iteration using FEMM simulation, we have found that the cusp magnetic field using six permanent magnets having the length of 1 m kept in the multi-pole cusp configuration at 60º each along the circumference of a 40 cm diameter cylinder full fill all the requirements of our physics studies. However to realize variable field free zone, variable cusp magnetic field geometry is necessary which can fulfilled by same number of electromagnets with profiled core material. Each electromagnet is made with double pan-cake windings using hollow copper pipes. The hollow copper pipes are used for forced water cooling and the pan-cake windings help for effective cooling using parallel lines. The physical dimensions of the rectangular magnet are in centimetre132, 19.5 and 14 respectively for length, width, and height. Similarly the physical dimension of vacoflux-50 core material dimensions are in centimetre 120, 2 and 12 respectively for length, width and height[23]. This core material is an alloy of iron and cobalt in the ratio 50:50. The physical property of core material i.e. relative permeability (> 4000), saturation flux density (> 2.0 Tesla) and Curie temperature (700 $^o$C) of Vacolfux-50 are too high and these properties are increased strength of VMMF.

To understand the scaling of magnetic field with respect to the magnet current, FEMM simulation has been carried out. The model for VMMF is constructed in FEMM tool, and the simulations are performed for air and for vacoflux-50 core (simply change the property of core material in the model) with core is magnetized by changing magnet currents ($I_{mag}$). The simulations are also performed with changing the shape of the edge surface of core in the model for showing the effect of core edge surface on magnetic field lines. The results obtained from the magnetic field simulations are described as follow.

Figure 2 shows the contour plot of the vacuum field lines in ($r$, $\theta$) plane of the device from FEMM simulation when core material is magnetized with 150A magnets current. It shows six electromagnets with embedded vacoflux-50 core are placed over a periphery of 40 cm diameter of the chamber. The colour code of magnetic field is same for all FEMM simulation figure and maximum magnetic field of figures is also less than the saturation magnetic field of core (2 Tesla). The magnetic field is measured over the different ($r$, $\theta$) plane of the device using triple axis gauss probe (F. W. Bell, model Z0A99-3202) and suitable gauss meter (F. W. Bell series 9900). The magnetic field is identical throughout all ($r$, $\theta$) planes of the device expect at the edge near magnet edge. The radial variation of the measured magnetic field ($B_m$) along the cusp region ($\theta = 0^o$) and the non-cusp region (the region exactly in between two magnet, ($\theta = 30^o$)) are shown in figure 3 for two different magnet currents ($I_{mag}$, 100A and 150A). It shows that the simulated magnetic field ($B_s$) matching well with measured value ($B_m$). Figure 4 shows contour plot of the vacuum field lines of an electromagnet with air as a core (figure 4(a)) and vacoflux-50 as a core (figure 4 (b)) when core is magnetized with 150A magnet current ($I_{mag}$). The magnetic field lines for air core are passing throughout the magnet but for the vacoflux-50, magnetic field lines are passing through only the core because of that magnetic field lines at the pole are highly dense for vacoflux-50 core. Thus magnetic field is high for vacoflux-50 core at the pole compared to that of air core. It is also taken care the edge of the core material which affects the magnetic field lines. The sharp edge of core material distorted the magnetic field lines not uniformly distributed over the surface of the core. To remove fringe due to sharp edge on magnetic field lines, the edge surface of core material is profiled as shown in figure 5(a). It shows the cross-sectional view of the vacoflux-50 material, the edge of vacoflux-50 is curved with a radius of curvature is 12.325 mm. Figure 5 (b) shows FEMM simulation of magnetic field lines at core edge surface is curved and figure 5 (c) shows the edge of the core surface is sharp (rectangular). It is clearly seen from figure 5 (b) that magnetic field lines are uniformly distributed over the surface of the edge for curved edge vacoflux-50 core but for sharp edge surface as shown in figure 5 (c) the field lines are not uniformly distributed and distorted at the edge surface and also the magnetic field is high at the edge corner. Thus results from figure 4 and 5, the magnetic field lines are profiled using profiled vacoflux-50 core.

Figure 6 shows the radial variation of (a) magnetic field and (b) its gradient along the cusp region as well as along the non-cusp region for air core and vacoflux-50 core when core is magnetized with magnet current ($I_{mag}$) 150A. It is clearly seen from both figures that at nearly field free region, gradient is weak and nearly same for both type of core but near the pole, the gradient in the magnetic field increases sharply for vacoflux-50 core compared to air core along the cusp region. Figure 6 (c) shows the variation of pole magnetic field ($B_p$ is field near the magnet at R = 20 cm) with respect to magnet current for air core and vacoflux-50 core. The rate of change of pole magnetic field ($B_p$) with respect to magnet current is very high 7.53 G/A for vacoflux-50 core compare to air core (2.15 G/A). Figure 6 (d) shows the radial variation of the magnetic field in the cusp region when magnets are energized with different currents. It is clearly seen that the radius of nearly field free region decreased with increasing magnet current. Thus, using electromagnets with the core material, field values are changed by changing magnet current without changing the number of the poles.

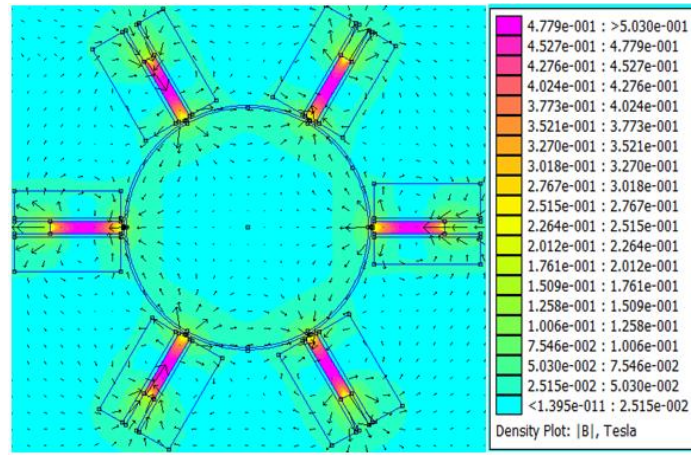

Figure 2.(a) Contour plot of the vacuum field lines in $(r,\theta)$ plane of MMF from FEMM simulation when core is magnetized with 150A, magnet current. Colour code for magnetic field is same for all FEMM simulation figures until and unless specified and maximum magnetic field is less than the saturation magnetic field of core.

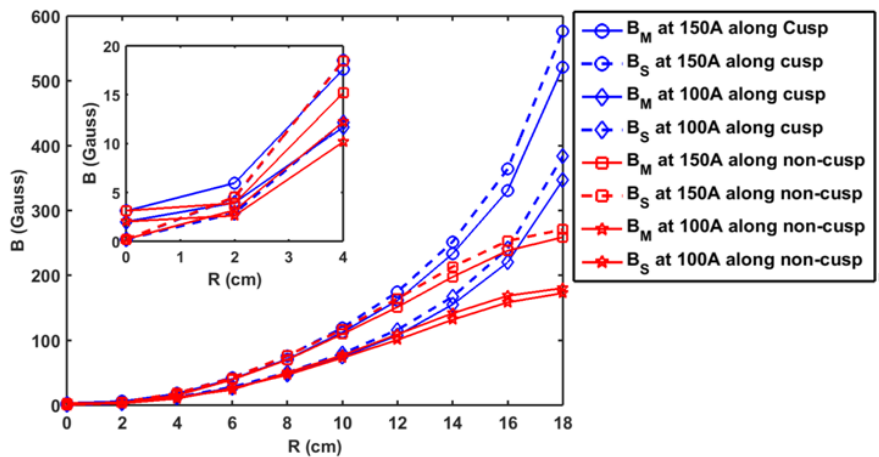

Figure 3. A comparison of the magnetic field between simulated ($B_s$), and measured ($B_M$) along the cusp and non-cusp regions when the core is magnetized with magnet currents 100 A, and 150A.

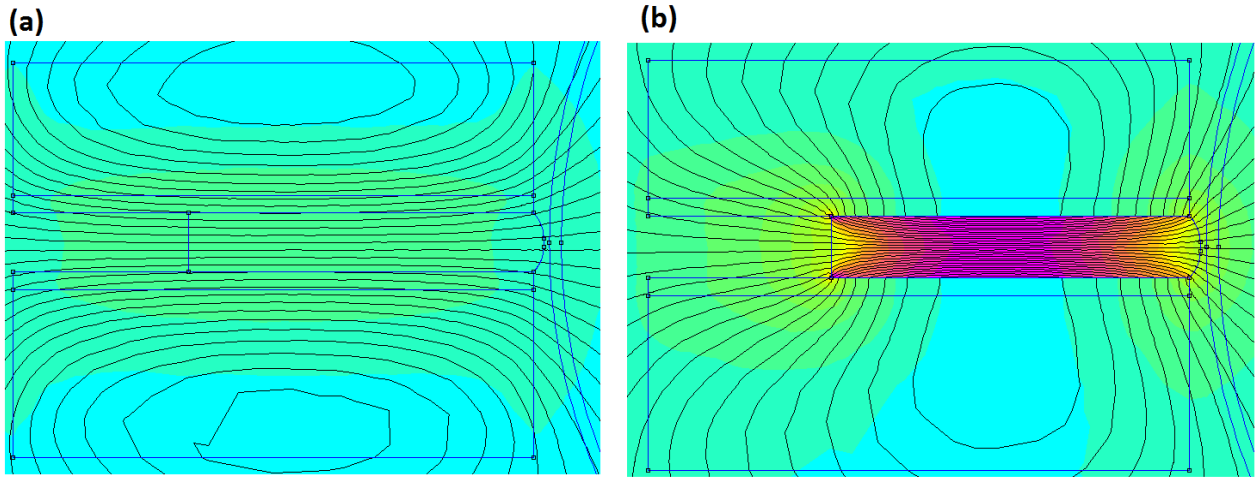

Figure 4: Contour plot of the vacuum field lines of electromagnet (a) for air core (b) for vacoflux-50 core when core is magnetized with magnet current 150A using FEMM simulation. Colour code is same as figure 1, and maximum magnetic field is less than the saturation magnetic field of core.

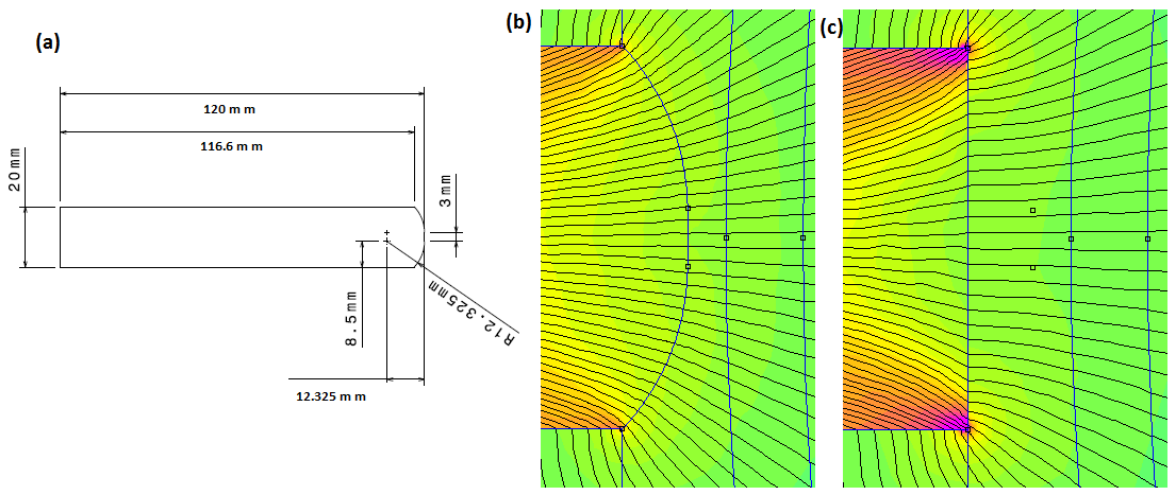

Figure 5. (a) Cross sectional view of vacoflux-50 showing curvature surface at one edge. Other two figure shows FEMM simulation of field lines for (b) edge of vacoflux-50 is curved and for (c) edge of vacoflux-50 is sharp (rectangular) when core is magnetized with 150A magnet current ($I_{mag}$). Colour code is same as figure 1, and maximum magnetic field is less than the saturation magnetic field of core.

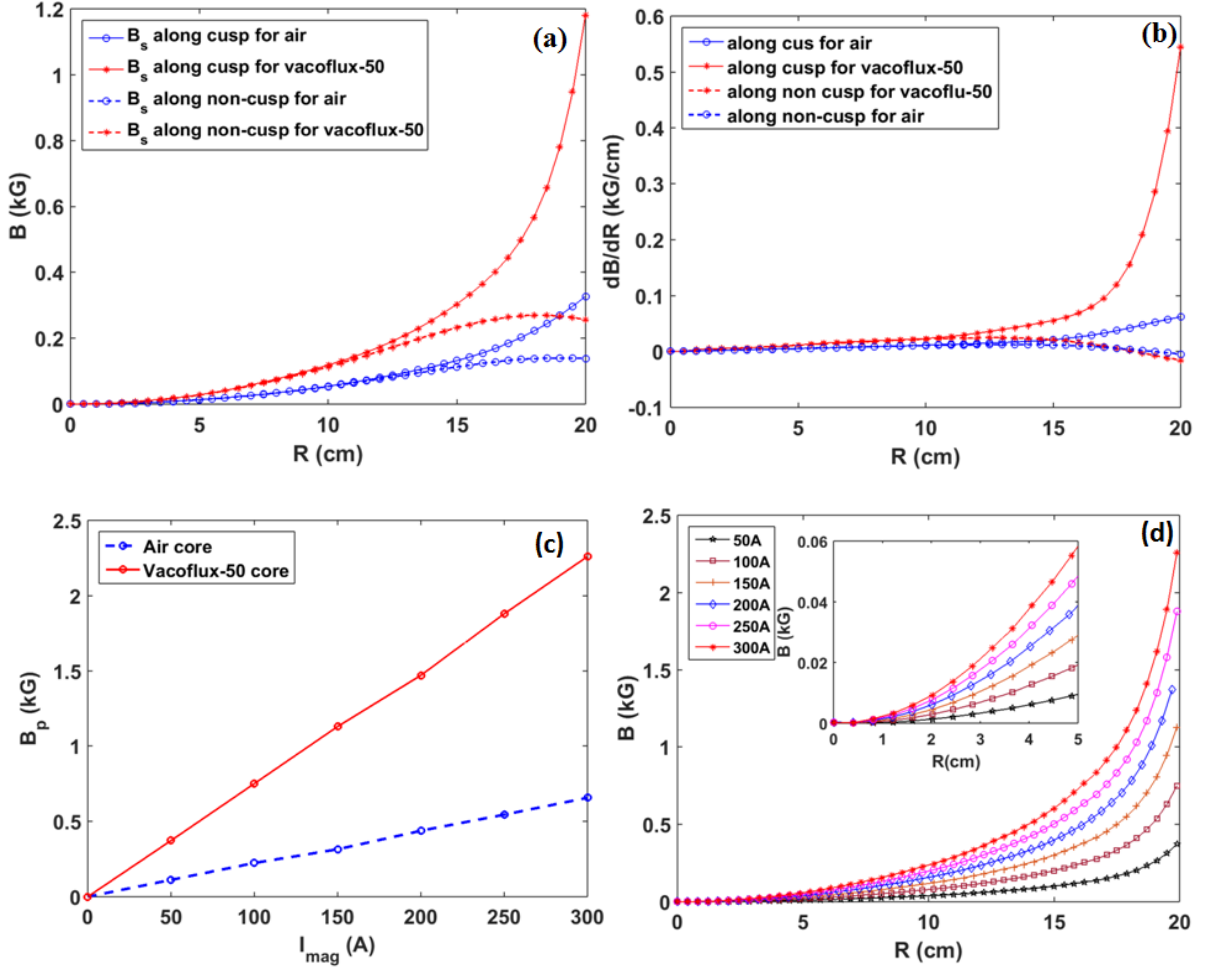

Figure 6. (a) Radial variation of magnetic field along cusp region and non-cusp region, (b) its gradient with respect to radial distance for vacoflux-50 and air core and magnet current ($I_{mag}$) is 150A. (c) Variation of pole magnetic field ($B_p$) (at R=20cm, near the pole of magnet) with magnet current for air Core and vacoflux-50 core.(d) The radial variation of magnetic field along the cusp region whenvacoflux-50 core is magnetized with different currents ($I_{mag}$).

## IV. Experimental results and Discussion

The basic plasma parameters such as electron temperature ($T_e$), plasma density ($n$), and floating potential ($V_f$) are measured using conventional Langmuir probe having length 5mm and diameter 0.25mm. Here figure 7 shows a typical Langmuir probe *V-I* characteristic at two regions of plasma by swiping bias voltage from -80V to +35V at two different radial locations when core is magnetized with 150A magnet current, and $2.0 \times 10^{-4}$ mbar argon gas background pressure. The MATLAB code is developed using technique used by Sheehan *et al.* [24], for analysis the *V-I* characteristic of plasma. Figure 8 is genetrated from the code and its show the procedure of analysis of electron temperature. A detailed analysis of electron temperature ($T_e$) has been reported in an earlier publication by the same authors [23].The trace-1 is *V-I* characteristic at the mid plane (($r, \theta$) plane, which is z = 65 cm apart from the filament) of the device and R= 0 cm. Similarly the trace-2 is *V-I* characteristic at R=14cm along the non-cusp region and same plane. Figure 8 (a) and (b) shows the log variation of electron current with respect to probe bias voltage. Here in figure 8 (a) and (b), trace-3 is log variation of plasma electron current and trace-4 is log variation of plasma cold electron current.

It is clearly observed from the two *V-I* characteristic that trace-1 (in figure 7) has highly negative floating potential ($V_f$), and constantly changing slope at the Maxwellian region of *V-I* characteristic, it is also seen from figure 8 (a) that cold electron current is away from the plasma electron current at Maxwellian region of *V-I* characteristic. Thus it is indicated that plasma at the centre is enriched of primary electron and it has two populations of electrons viz. hot electron and cold electron and its nature is bi-Maxwellian because of cusp magnetic field confined the primary ionizing electron in field free region. But in trace-2, $V_f$ is near to zero and slope of *V-I* is sharper in Maxwellian region of *V-I* characteristic, it is also seen from figure 8 (b) that cold electron current and plasma electron current matches at Maxwellian region of *V-I* characteristic, thus plasma at R = 14 cm has very low population of primary electrons and its nature is Maxwellian. Since, the source is located in null region and cusp magnetic field confines the primary electrons emitted from the source. As we move radially outward across the magnetic field, the magnetic field increases after null region and primary electrons follow the magnetic field lines, and move back and forth between two pole and makes collision with background argon gas atoms thus primary electron cannot diffuse too much distance across the magnetic field and confined in null region of cusp magnetic field. Hence cusp magnetic field acts like a filter for primary electrons across the magnetic field.

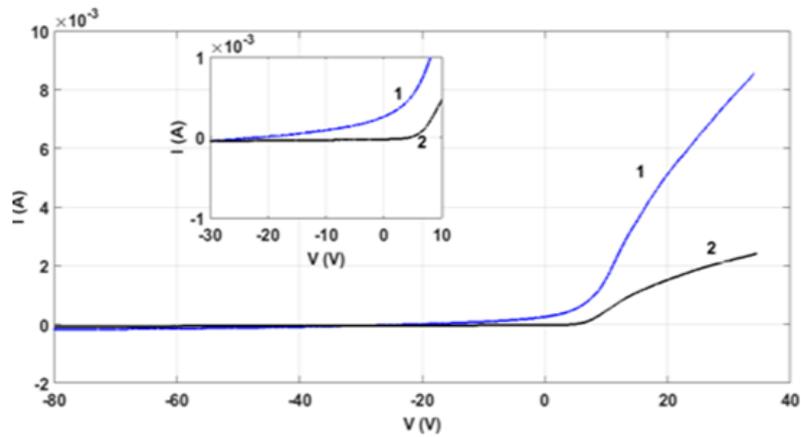

Figure 7: A typical Langmuir probe *V-I* characteristic of plasma at two radial locations R=0cm and R=14cm along the non-cusp region when core is magnetized with magnet current 150 A.

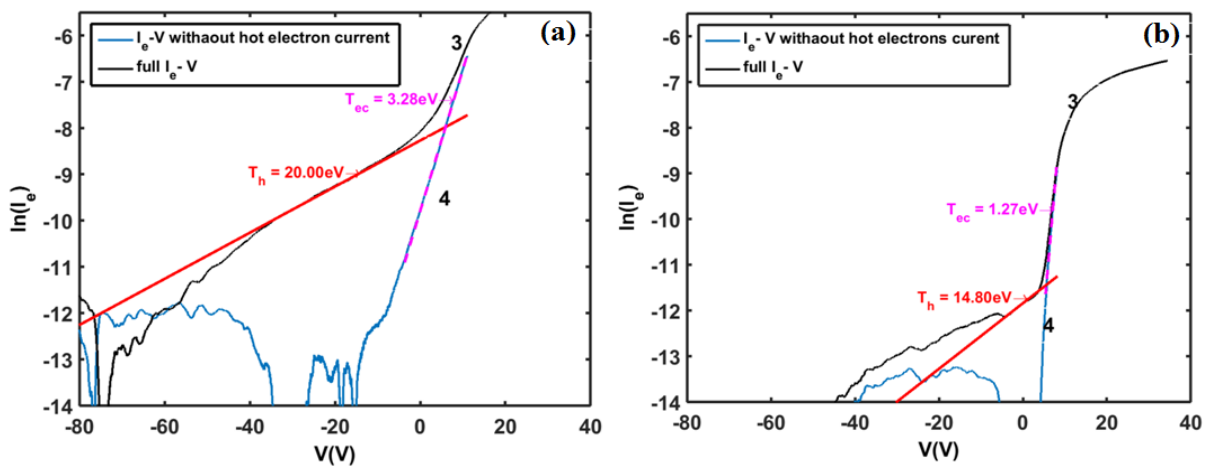

Figure 8. Plots for the variation of $\ln(I_e)$ with bias potential(*V*),(a) at R=0cm and (b) at R=14cm along the non-cusp region when core is magnetized with magnet current 150A.

### a. Analysis of electron temperature ($T_e$) and plasma density

In general, bi-Maxwellian plasmas consist of two types of electrons with hot electron temperature component($T_h$) and cold electron temperature component ($T_{ec}$) [24-25]. Kinetic definition of Effective temperature($T_e$) can be expressed as:

$$T_e = \frac{1}{3} m_e \int_{-\infty}^{\infty} v^2 f(v) dv \quad \text{...............(1)}$$

An effective temperature can be derived by assuming electron energy distribution function (EEDF) as bi-Maxwellian, $f(v) = \alpha f_h(v) + (1-\alpha) f_c(v)$, Where $f_h(v)$ is the hot electron distribution function and $f_c(v)$ is the cold electron distribution function. $\alpha$ is the ratio of hot electron current to electron current at plasma potential [24-25]. Hence, for bi-Maxwellian plasma, the effective temperature can be derived as [24]:

$$T_e = (1-\alpha) T_{ec} + \alpha T_h \quad \text{...........................(2)}$$

A detailed analysis of electron temperature ($T_e$) has been reported in our earlier publication [23].

To measure the plasma density Ion saturation current ($I_{isat}$) is considered. In a collision less single ion species plasma the ion saturation current to the probe is expressed [26-27] by

$$I_{isat} = \gamma e \, n \, A_P \sqrt{\frac{k_B T_e}{m_i}} \quad \text{.........................(3)}$$

Where $n$ is the plasma density, $A_p$ is the probe area, $m_i$ is the mass of the ion. The constant $\gamma$ depends on the probe sheath thickness. For, $\xi = r_p / \lambda_D > 3$, where $r_p$ is probe radius and $\lambda_D$ is Debye length,

$$\gamma = \zeta \chi^\beta / 4 \quad \text{...............................(4)}$$

Where $\chi = \frac{e(V_s - V_p)}{k_B T_h}$ and $\zeta = 1.37 \, \xi^{0.06}$ and $\beta = \xi^{0.52}$ [27].

The equation of $I_{isat}$ for $\xi = r_p / \lambda_D \geq 3$ is given by orbital motion limited (OML) theory [28],

$$I_{isat} = 1.13 e \, n \, A_P \, \chi^{0.5} \sqrt{\frac{k_B T_e}{2 \pi m_i}} \quad \text{....................(5)}$$

The density can be examined by eq. (3). The values of $T_e$ and $n$ give us the information of cusp characteristics in term of leak width. We now describe the experimental results on confinement property of VMMF i.e. leak width of plasma and its effect on confinement of primary electrons, plasma density, particle confinement time, density fluctuation with changing magnetic field.

### b. Leak width

The leak width (*d*) of quasi-neutral plasma flowing out of a cusp and cross field diffusion is dominated by Bohm diffusion is given by [12-13]

$$d = \left( \frac{D_\perp L}{C_s} \right) \quad \text{........................(1)}$$

where, $D_\perp$ is diffusion coefficient across the magnetic field near the pole of magnet, $L$ is scale length of magnetic field (*B*) and $c_s$ is the ion acoustic speed.

In low background gas pressure, R. Jones (1981) relates the leak width (*d*) to plasma turbulence [19]. Micro-instabilities which are unstable in a cusp sheath (e.g. drift wave, ion cyclotron waves) give rise to Bohm like diffusion [18, 19]. The relationship between the relative electron density fluctuation level $dn_e/n_e$ and the leak width (*d*) for turbulent line cusp [19],

$$d \geq \frac{dn_e}{n_e} \frac{T_e}{T_i} r_{Li} \quad \ldots\ldots\ldots\ldots\ldots (2)$$

Where $n_e$ the electron density is $r_{Li}$ is ion Larmor radius, $T_e$ and $T_i$ are electron and ion temperature respectively.

The density fluctuation have been measured at R=16cm near the pole in cusp region. The power spectra of density fluctuation at different magnet current are shown in figure 9 (a). The observed turbulence have broad band spectra with significant power. Hence considering turbulant line cusp, we use equation (2) for leak width calculation. In equation 2, we consider $\frac{dn_e}{n_e} \approx \frac{\delta I_{esat}}{I_{esat}}$ where $I_{esat}$, the electron saturation current is observed in the cusp region at R=16cm and it is also considered that the ion temperature ($T_i$) is 10 times less than the electron temperature ($T_e$). The electron temperatures ($T_e$) and electron saturation currents ($I_{esat}$) are measured from same single Langmuir probe at R=16cm near the pole of magnet. The leak width (*d*) from equation (2) at different magnet currents are calculated and compared with leak width calculated from hybrid gyro-radius $d=2(r_i r_e)^{0.5}$ [2]. The leak widths are good match at high magnetic field values but not for low magnetic field as shown in figure 9 (b) and these results are consistent with results obtained by Bosh *et al.* [12-13]. The leak width of plasma is decreased with increasing magnet current and hence decreasing the loss rate of bulk plasma. Thus it effect the plasma parameters i.e. plasma density, floating potential, and particle confinement time.

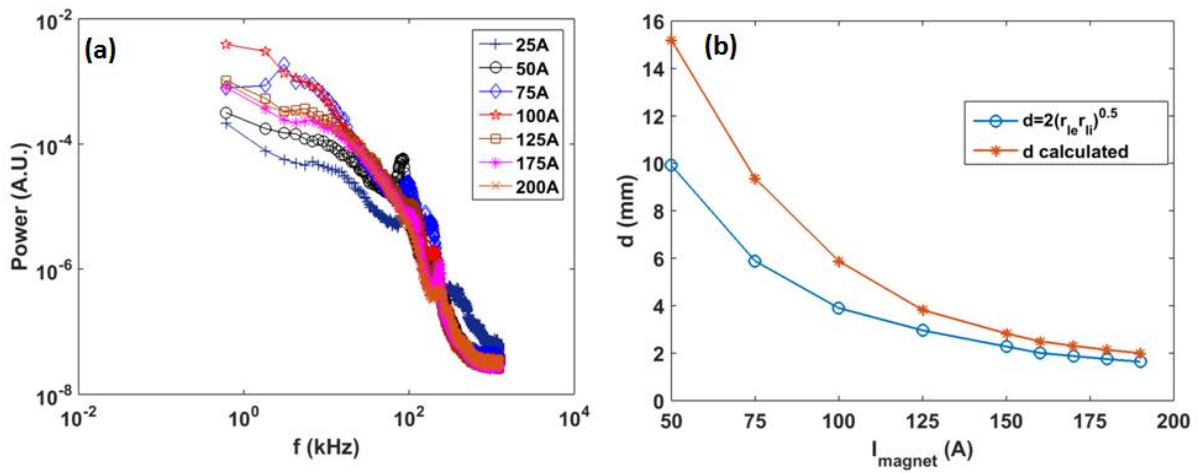

Figure 9.(a) Density autopower spectra at R=16cm in cusp region near the pole and (b) comparision of leak width between calculated from equation (2) and calculated from hybrid gyro radius at R=16 cm in cusp region when core is magnetized with different current and $2.0 \times 10^{-4}$ mbar argon gas background pressure.

The device has facilities to change the cusp magnetic field values by changing the magnet current. The following experimental results are obtained when core is magnetized with different magnet currents. The radial variation of plasma parameters (floating potential, plasma densities,and electron temperature) are measured across the magnetic field along the non-cusp region (the region in between the two consecutive magnets) and at the mid (*r*, *θ*) plane (which is z=65cm apart from the filament) of the device, the argon gas pressure is $2 \times 10^{-4}$ mbar. The schematic diagram of this plane is shown in figure 10 (a). Figure 10 (b) shows the photograph of argon plasma taken from the viewport fitted at end of the chamber in multi cusp plasma device.The bulk plasma confined in cusp magnetic field is diffused across the magnetic field and stream out along the field lines thus the magnetic field control the radial variation of mean plasma parameters. Now we described the effect of magnetic field (leak width) on radial variation of plasma parameter with core is magnetized with different magnet current.

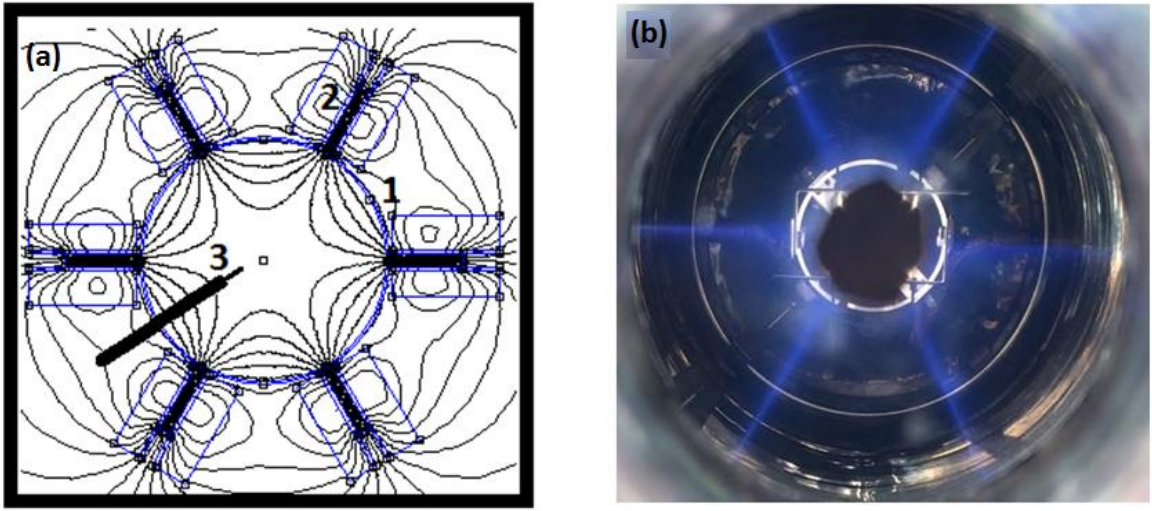

Figure 10: (a) Schematic diagram of mid plane ((r, θ) z = 65 cm from the filament) of the device. 1: Chamber cross section, 2: Magnet, 3: Langmuir probe across the magnetic field along the non-cusp region (exactly in between two consecutive magnets). (b) Photograph of argon plasma taken from the viewport fitted at end flange of the chamber.

Figure 11 (a) shows the radial variation of floating potential across the magnetic field along the non-cusp region. The cusp magnetic field confined the primary electrons in confined region (null region) and scavenged these electron across the magnetic field thus floating potential is high negative at confined region and decreased radially outwards across the magnetic field. The floating potential of confined region also become more and more negative with increasing magnet current because confinement of primary electrons is increased with increasing magnet current. The scavenging radial length across the magnetic field (i.e. floating potential become nearly -5 V at radial distance 7 cm, 6 cm, and 5 cm for 100 A, 150 A, and 200 A magnet current respectively) decreased with increasing magnet current. Thus VMMF scavenged the primary electrons across the magnetic field and act as like filter, and it has also control over filtering strength (scavenging length). The confined primary electron at confined region move back and forth between two poles and make collision with background gas atoms, the collision may ionize or excite the argon gas atom. In figure 10 (b) the deep purple colors are first excitation energy of argon atom and its tack place only when the temperature of the electron is more than 3 eV thus deep purple color shows the path of primary electrons.

The figure 11 (b) and (c) show the radial variation of plasma density ($n$) and electron temperature ($T_e$) along the non-cusp region when the core is magnetized with 100 A, 150 A, and 200 A magnet current. The plasma is dense and density is increased with increasing magnet current in a confined region. The radial uniformity of plasma density across the magnetic field along the non-cusp region is 4 cm, 5 cm and 6 cm for 100 A, 150 A and 200 A magnet current respectively. As we increase the magnetic field the leak width of plasma decreases thus parallel diffusion lose rate decreased. Beside plasma density increases with decreasing leak width thus plasma species make more collisions with each other. As a result perpendicular diffusion coefficient increases and hence radial plasma density uniformity are increased with magnetic field (magnet current) [29]. The confined plasma is hot(electron temperature is high up to 5 cm radial distance) as shown in figure 11 (b), because cusp magnetic field confined primary electron in confined region and make more collision with background gas atoms and losing its energy. As we move radial outward across the magnetic field the magnetic field scavenged the primary electrons because of that electrons thus temperature decreased across the magnetic field. We also notice that the electron temperature decreased with increasing the magnet current in confined region. As we increase the magnet current (magnetic field) the leak width (loss rate) of plasma decreases thus plasma density increased in confined region and plasma species spends more time to losing its energy by collision with each other as a result electron temperature decreases with magnetic field.

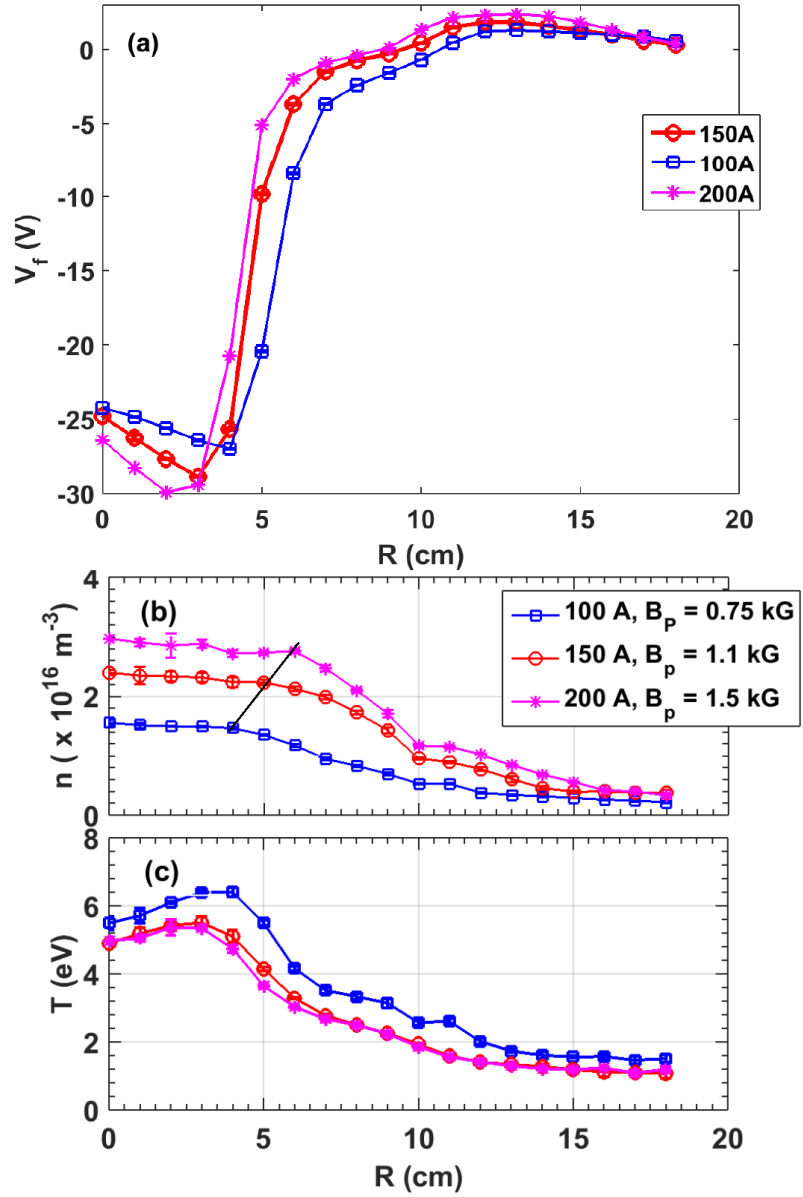

Figure 11: The Radial variation of mean plasma parameters: (a) floating potential (b) plasma density ($n$) and (c) electron temperature ($T_e$), across the magnetic field and along non-cusp region, when core is magnetized with three different magnet currents.

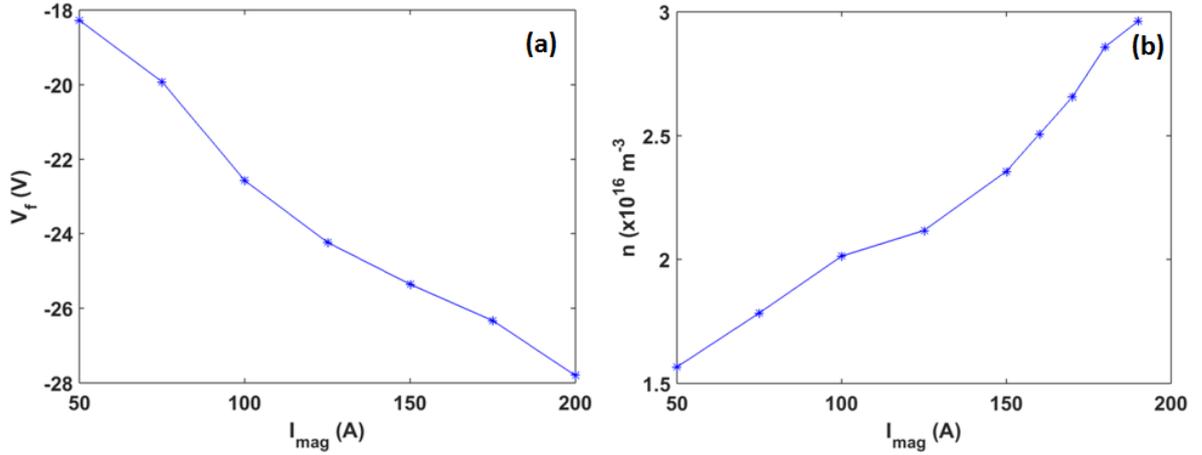

Figure 12: Variation of (a) floating potential ($V_f$) and (b) plasma density ($n$) at R = 0 cm with different magnet ($I_{mag}$) current.

Figure 12 (a) shows the variation floating potential at R = 0 cm with magnets cureent ($I_{mag}$).The folating potetial is very high and negative at R = 0 cm beacuse cusp magnetic field confined the primary electron as we discussed earlier.The leak width of plasma is decreased with increasing magnet current as shwon in figure 8 (b) because of that leak rate of plasma decreased hence confinement of primary electron increased. Thus, the floating potential becomes more and more negative with increasing the magnet current as shwon in figure 12 (a).These confined primary electrons move back and forth between two poles of magnets and makes more collision with neutral background argon gas atoms and enhance the plasma density as well as plasma leak rate of plasma also decreased with increasing magnet current (magnetic field) thus the plasma density in confined region (R = 0 cm) is increased with increasing magnet current as shown in figure 12 (b).

The particle confinement time is measured from time profile of ion saturation current in afterglow plasma. The afterglow plasmas can be produced by switching off the biasing voltage of the multi-filamentary discharge. The discharge power supply is switched oFF and ON for 500 ms and simultaneously measure the ion saturation current across the 10 kΩ resistance using simple Langmuir probe and applied voltage at probe tip is -100 V. Figure 13 (a) shows the time profile of normalizes ion saturation current ($I_{isat}$) when core is magnetized with 150 A magnet current and probe is located at R = 0 cm in afterglow plasma. The time is taken for ion saturation current to reduces 1/e of its initial value is called the particle confinement time. From figure 13 (a), it can be observed that time profile of ion saturation current has two exponential regions thus plasma has two particle confinement time. As we discussed earlier cusp magnetic field confines primary electrons, after switch offthe discharge power supply the confined energetic (primary) electron move back and forth motion along the field lines and ionize the background gas. First rapid exponential is due to plasma electron and second exponential due to energetic trapped electrons in cusp magnetic field [29]. Figure 13 (b) shows the variation of particle confinement time when the core is magnetized with a different current. As we discussed earlier that the leak width of plasma decreased with increasing magnet current (or magnetic field), because of that plasma density at confined region is increased and leak rate of plasma decreased, thus particle confinement time is increased with magnet current (or magnetic field) as shown in figure 13 (b). The stability of plasma also increased with increasing the particle confinement time and plasma density. The increase in plasma stability also seen from figure 13 (c) that fluctuation in ion saturation current (density) at R = 0 cm is decreased with increasing magnet current. Figure 13 (d) shows the frequency spectra of these density fluctuations. From the frequency power spectrums (figure 9 (a) and 13 (d)), the fluctuation observed at the cusp region is not present at null region. Thus the instability observed at the cusp region is not propagating at the confined region. The detailed analysis of instbilitis at both region required more studies and will be publised in next article.

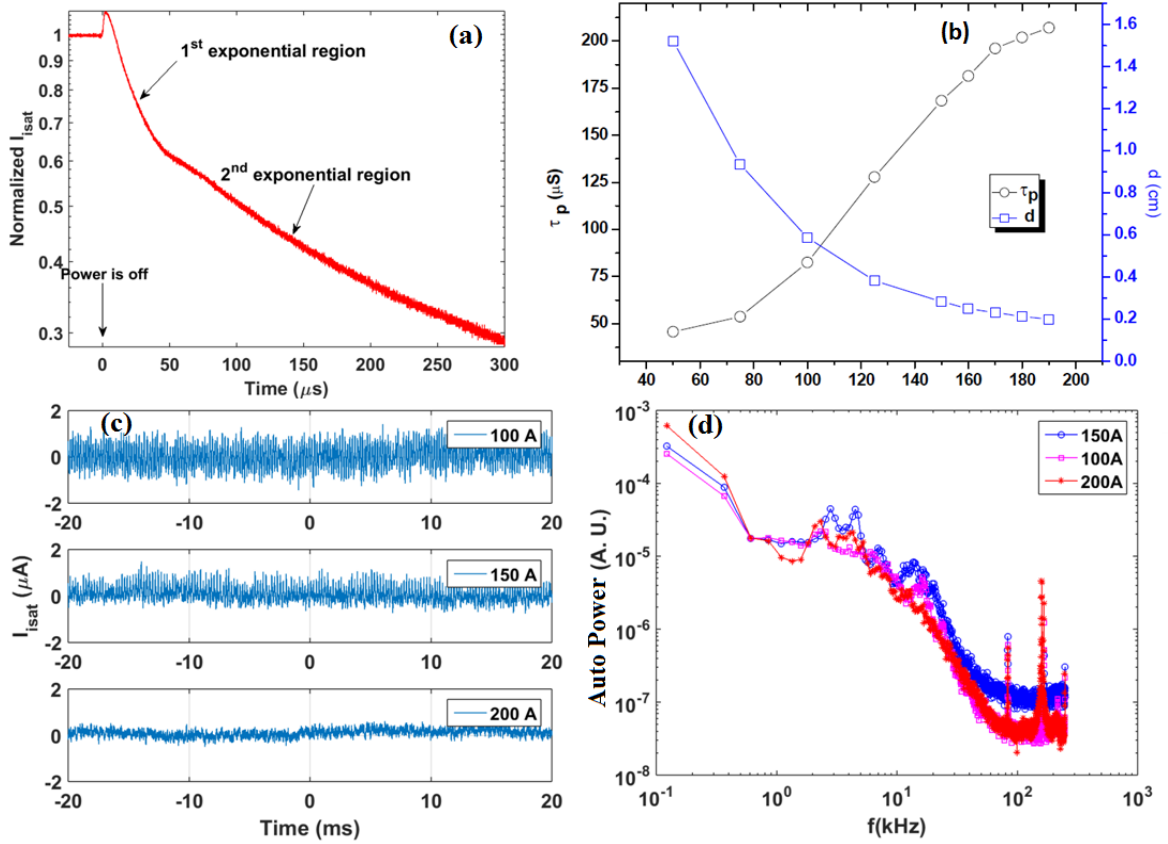

Figure 13: (a) Time profile of normalized ion saturation current ($I_{isat}$) when core is magneized with 150A magnet current in after glow plasma. (b) Variation of particle confinement time ($\tau_p$) with magnet current ($I_{mag}$) when core is magnetized with differet magnet current ($I_{mag}$).(c) Time profile of ion saturation current and (d) frequency power spectra of density fluctuation at R=0cm when core material is magnetized with different magnet current ($I_{mag}$).

## V. Conclusion

A variable multi-pole line cusp magnetic field is generated over large cylindrical volume (1 m axial length and 40 cm diameter of the chamber) and similar field is demonstrated using FEMM simulation. The profiled vacoflux-50 core material is profiled the magnetic field lines and sharp edge effect of the core on magnetic field lines are removed by profiling core. The magnetic field simulation results show generated VMMF has full control over a nearly field free region (null region) without changing poles of magnets and the rate of change of pole magnetic field with respect to magnet current is very high for vacoflux-50 core compared to simple air core. The argon plasma has been characterized in VMMF. The experimental result shows the cusp magnetic field confines the primary electron and plasma has bi-Maxwellian nature in the confined region. The confinement of plasma parameter (primary electron, plasma density and particle confinement time) improved with increasing magnetic field because of plasma leak width (leak rate) decreased thus stability of plasma increased with the magnetic field. The cusp magnetic field also filters the confined primary electrons across the magnetic field and the filtering strength of primary electrons across the magnetic field is increased with the magnetic field. The control over the null region of MMF, plasma density, radial uniformity of plasma density with changing magnetic field and without changing the number of the poles of magnets are an essential property of new generation particle accelerator, plasma sources as well as NBI system of Tokamak.

**Acknowledgement:**

The authors would like to express sincere gratitude to Dr. Mainak Bandopadhyay for giving valuable suggestions for improving the manuscript as well as reviewing the manuscript. The authors would like to acknowledge Dr. Joydeeep Ghosh for numerous fruitful discussions. The author A. D. P also acknowledges Arghya Mukherjee for correcting the manuscript.

**Acknowledgement:**

The authors would like to express sincere gratitude to Dr. Mainak Bandopadhyay for giving valuable suggestions for improving the manuscript as well as reviewing the manuscript. The authors would like to acknowledge Dr. Joydeeep Ghosh for numerous fruitful discussions. The author A. D. P also acknowledges Arghya Mukherjee for correcting the manuscript.

**Reference:**

1. R. Limpaecher, and K. R. MacKenzie, Rev. Sci. Instrum., **44**, 726 (1973).
2. M.G. Haines, Nucl. Fusion **17**, 811 (1977).
3. I. Spalding, Advances *in plasma physics*, edited by A. Simon and W. B. Thomson (Interscience, New Yourk, **4**, 1971)79.
4. L. R. Grishma *et. al.*, Fusion Engineering and Design, **87,** 1805 (2012).
5. W. L. Striling, P. M. Ryan, C. C. Tsai, and K. N. Leung, Rev. Sci. Instrum.,**50**, 102 (1979).
6. M. Sugawara, *Plasma Etching Fundamentals and Applications*, Oxford University Press, Oxford, England, (1998).
7. R. Günzel, E. Wieser, E. Richter, J. Steffen, Journal of Vacuum Science and Technology B, **12,** 927 (1994).
8. M. A. Lieberman, A. J. Lichtenberg, *Principles of Plasma Discharges and Material Processing*, Wiley–Interscience, New York, (1994).
9. S. Mukherjee and P. I. John, Sur. and Coatings Technol., **93,**188 (1997).
10. M. Hosseinzadeh and H. Afarideh, Nucl. Instrum. And Meth. **A,** 735**,** 416 (2014).
11. K. N. Leung, T. K. Samec, and A. Lamm, Phys. Lett. A, **51,** 490 (1975).
12. R. A. Bosch, and R. L. Merlino, Phys. Fluids, **29(6),** 1998 (1986).
13. R. A. Bosch, and R. M. Gilgenbach, Phys. Lett. A, **128 (8),** 437 (1988).
14. N. Herkowitz, K.N. Leung, and K. R. MacKenzie, Phys. Fluids, **19**, 1045 (1976).
15. R. Jones, Nuovo Cimento, **34**, 157 (1982).
16. H. Kozima, S. Kawamoto, and K. Yamagiwa, Phys. Lett., **86**, 373 (1981).
17. R. Jones, Plasma Phys., **21**, 505 (1979).
18. R. Jones, Plasma Phys., **23**, 381 (1981).
19. N. Herkowitz, J. R. Smith, and H. Kozima, Phys. Fluids, **22**, 122 (1979).
20. Earl R. Ault and K. R. MacKenzie, Rev. Sci. Instrum., **44**, 1697 (1973).
21. J. H. Kim, Phys. Procedia, **66**, 498 (2015).
22. D. Meeker, *"Finite element method magnet"*, Version 4.2, Users Manual 2015, http://www.femm.info.
23. A. D. Patel, M. Sharma, N. Ramasubramanian, R. Ganesh, and P. K. Chattopadhyay, Rev. Sci. Instrum., **89**, 043510 (2018).
24. M. U. Siddiqui, and N.Hershkowitz, Phys. Plasma, **21**, 020707 (2014).
25. F. F. Chen, Plasma Sources Sci. Technol., **18,** 035012 (2009).
26. Sayak Bose, Manjit Kaur, P. K. Chattopadyay, J. Ghosh, Y. C. Saxena, and R. Pal, J. Plasma Phys., **83,** 615830201 (2017).
27. Isaac D. Sudit, and R. Claude Woods, J. Appl. Phys., **76**, 4488 (1994).
28. C. M. Cooper, D. B. Weisberg, I. Khalzov, J. Milhone, K. Flanagan, E. Peterson, C. Wahl, and C. B. Forest, Phys. Plasmas, **23**, 102505 (2016).
29. S. Aihara, M. Fujiwara, M. Hosokawa, and H. Ikegami, Nucl. Fusion, **12**, 45 (1972).